\documentclass{emulateapj}

\begin{document}

\shortauthors{Luhman \& Esplin}
\shorttitle{SED of the Coldest Known Brown Dwarf}

\title{The Spectral Energy Distribution of the Coldest Known
Brown Dwarf\altaffilmark{1}}

\author{K. L. Luhman\altaffilmark{2,3} and T. L. Esplin\altaffilmark{2}}

\altaffiltext{1}
{Based on observations made with the {\it Spitzer Space Telescope},
the NASA/ESA {\it Hubble Space Telescope}, Gemini Observatory, and
the ESO Telescopes at Paranal Observatory.}

\altaffiltext{2}{Department of Astronomy and Astrophysics,
The Pennsylvania State University, University Park, PA 16802, USA;
kluhman@astro.psu.edu}
\altaffiltext{3}{Center for Exoplanets and Habitable Worlds, The 
Pennsylvania State University, University Park, PA 16802, USA}

\begin{abstract}

WISE J085510.83$-$071442.5 (hereafter WISE 0855$-$0714)
is the coldest known brown dwarf
($\sim250$~K) and the fourth closest known system to the Sun (2.2~pc).
It has been previously detected only in the $J$ band and two mid-IR bands.
To better measure its spectral energy distribution (SED), we have performed
deep imaging of WISE 0855$-$0714 in six optical and near-IR bands
with Gemini Observatory, the Very Large Telescope, and the
{\it Hubble Space Telescope}. 
Five of the bands show detections, although one detection is marginal 
(S/N$\sim$3). 
We also have obtained two epochs of images with the {\it Spitzer Space
Telescope} for use in refining the parallax of the brown dwarf.
By combining astrometry from this work and previous studies, we have
derived a parallax of $0.449\pm0.008\arcsec$ (2.23$\pm$0.04~pc).
We have compared our photometry for WISE 0855$-$0714 to data
for known Y dwarfs and to the predictions of three suites of models
by \citet{sau12} and \citet{mor12,mor14} that are
defined by the presence or absence of clouds and non-equilibrium chemistry.
Our estimates of $Y-J$ and $J-H$ for WISE 0855$-$0714 are redder than colors
of other Y dwarfs, confirming a predicted reversal of near-IR colors to
redder values at temperatures below 300--400~K.
In color-magnitude diagrams, no single suite of models provides a clearly
superior match to the sequence formed by WISE 0855$-$0714 and other Y dwarfs.
Instead, the best fitting model changes from one diagram to the next.
Similarly, all of the models have substantial differences from the
SED of WISE 0855$-$0714.
As a result, we are currently unable to constrain the presence of clouds or
non-equilibrium chemistry in its atmosphere.

\end{abstract}


\section{Introduction}

With the discovery of the brown dwarf Gl~229~B
\citep{nak95,opp95}, it became possible to begin testing atmospheric models
within the temperature gap of 100--2000~K between the Jovian planets and
low-mass stars \citep{all96,mar96}.
Large variations in atmospheric properties (e.g., gas abundances,
opacity sources, presence and composition of clouds) are expected across
that temperature range \citep{bur01,mar15}.
As a result, brown dwarfs have been sought at progressively 
lower temperatures to more fully test models of cool atmospheres.
Most of the recent progress in surveys for the coldest brown dwarfs
has been enabled by the {\it Wide-field Infrared Survey Explorer}
\citep[{\it WISE},][]{wri10}, which obtained mid-infrared (IR) images of
the entire sky.
Those data have been used to greatly expand samples of late-T dwarfs
(500--800~K) and to discover members of a new, cooler spectral class,
the Y dwarfs \citep[$<500$~K,][]{cus11,cus14,kir12,kir13,tin12,pin14,sch15}.
A few additional brown dwarfs that are likely members of the Y class
have been found as companions through near-IR adaptive optics imaging
\citep{liu11,liu12} and mid-IR imaging \citep{luh11} with the
{\it Spitzer Space Telescope} \citep{wer04}.
Because the optical and near-IR fluxes of brown dwarfs decrease rapidly
with lower temperatures at spectral types later than T6
\citep[$<$1000~K,][]{cus11,kir11},
it is challenging to measure the spectral energy distributions (SEDs) 
of Y dwarfs for comparison to the predictions of atmospheric models.

The coldest known brown dwarf is WISE J085510.83$-$071442.5 (hereafter
WISE 0855$-$0714), which has a temperature of $\sim$250~K based on 
its absolute magnitude at 4.5~\micron\ \citep{luh14b}.
It has a distance of slightly more than 2~pc, which makes it the fourth
closest known system to the Sun.
Because of the combination of its close proximity and low temperature,
WISE 0855$-$0714 is a uniquely appealing target for studies of cool
atmospheres.
In \citet{luh14b}, the detections of WISE 0855$-$0714 
were limited to broad-band filters at 3--4 and 4--5~\micron\ from
{\it WISE} and {\it Spitzer}. It was not detected in moderately deep
near-IR images ($J=23$). Additional limits were subsequently placed on its
flux in the $z\arcmin$, $Y$, and $H$ bands \citep{bea14,kop14,wri14}.
\citet{fah14} reported a 2.6~$\sigma$ detection ($J\sim25$) in a custom filter
within the $J$ band, and they interpreted the resulting color
between $J$ and 4.5~\micron\ as evidence of water ice clouds based on
a comparison to models of brown dwarfs with and without clouds 
\citep{sau12,mor12,mor14}.
However, \citet{luh14c} found that those data were best matched
by cloudless models that employed non-equilibrium chemistry.

Deeper images of WISE 0855$-$0714 in multiple bands are needed 
to better constrain its SED. Because of its exceptional sensitivity,
the {\it Hubble Space Telescope} ($HST$) has been used for much of the
near-IR imaging and spectroscopy of Y dwarfs
\citep{cus11,cus14,kir12,kir13,luh14wd,sch15}.
In this paper, we present $HST$ observations of WISE 0855$-$0714 that we have
performed in F850LP ($\approx z\arcmin$), F105W (wide $Y$), F110W (wide $Y+J$),
and F127M (narrow $J$).
We also have obtained images in the $i\arcmin$ band with Gemini Observatory
and in a filter encompassing the $H$-band continuum of brown dwarfs with the
Very Large Telescope (VLT). Finally, we have observed WISE 0855$-$0714 at
additional epochs with {\it Spitzer} to refine its parallax.

\section{Observations}

In Table~\ref{tab:log}, we have compiled the telescopes, instruments,
filters, exposure times, dates, and program identifications for
our imaging of WISE 0855$-$0714. We have also included all previous
{\it Spitzer} observations of the object \citep{luh14b,luh14c,mel15}
and two photometric monitoring campaigns with {\it Spitzer} that will be
presented by T. Esplin (in preparation).

\subsection{Selection of Filters}

\begin{figure}
\epsscale{1}
\plotone{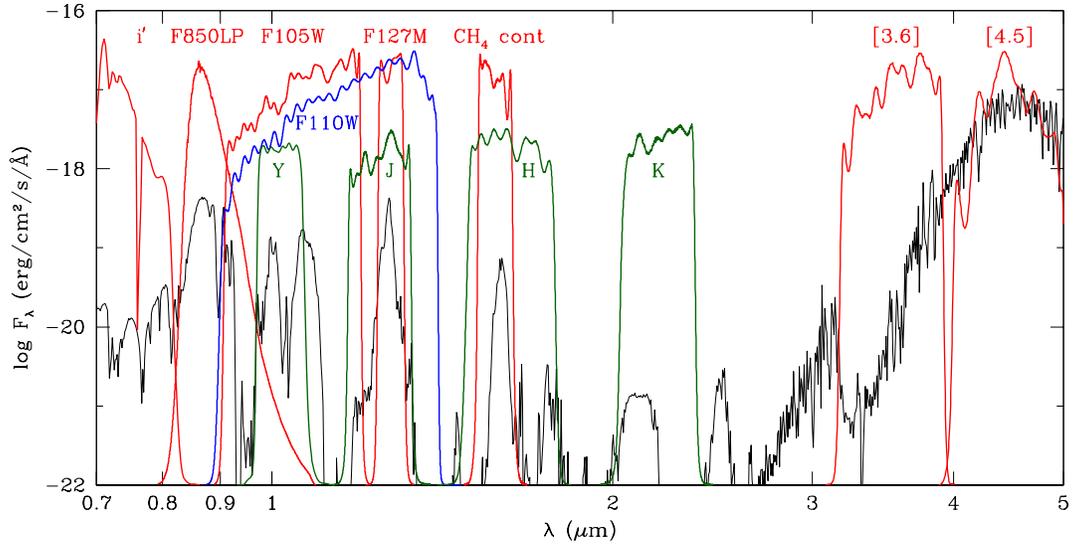}
\caption{
Example theoretical spectrum of a cold brown dwarf
\citep[250~K at 2~pc,][]{mor14} compared to the transmission profiles of the
bands in which we have obtained images of WISE~0855$-$0714 (red and blue).
For reference, the profiles for $Y$, $J$, $H$, and $K$ (MKO) have
been included (green). The profiles are plotted on a linear scale with
arbitrary normalizations.
}
\label{fig:filters}
\end{figure}

We have observed WISE 0855$-$0714 in six optical and near-IR filters
to better characterize its SED and in a mid-IR band at 4.5~\micron\ with
{\it Spitzer} to provide additional astrometry for refining its parallax.
We also have measured photometry from previous images in a second
{\it Spitzer} band at 3.6~\micron.
To illustrate how the bandpasses of these filters align with the
expected spectral features of WISE 0855$-$0714, 
we plot in Figure~\ref{fig:filters} the filter transmission profiles
and an example of a model spectrum for a 250~K brown dwarf at a distance
of 2~pc \citep{mor14}. When possible, we selected filters that
primarily encompassed continuum emission rather than deep absorption bands,
as in the case of F127M on $HST$ and the CH$_4$ continuum filter on VLT.
For reference, we include in Figure~\ref{fig:filters} the
$YJHK$ filters from the Mauna Kea Observatories photometric system
\citep{tok02}.

\subsection{{\it Spitzer} IRAC Images}
\label{sec:irac}

WISE 0855$-$0714 has been observed on multiple occasions by
{\it Spitzer}'s Infrared Array Camera \citep[IRAC;][]{faz04}.
It contains two $256\times256$ arrays that are currently operating,
each with a plate scale of $1\farcs2$~pixel$^{-1}$ and a field of view
of $5\farcm2\times5\farcm2$. The two arrays cover adjacent areas on the sky
in different filters that are centered at 3.6 and 4.5~\micron\ (denoted as
[3.6] and [4.5]). Point sources in these bands exhibit FWHM$=1\farcs7$.

All IRAC observations of WISE 0855$-$0714 are listed in Table~\ref{tab:log}.
\citet{luh14b} obtained the data on the first two dates, which were
in 2013 and 2014. On the first date, WISE 0855$-$0714 was imaged
in both [3.6] and [4.5]. The primary purpose of the second observation
was the measurement of astrometry for WISE 0855$-$0714, so it was
observed only in the band in which it is brightest, namely [4.5]. 
The next two epochs of data were collected in 2014 by \citet{luh14c},
which were used to measure additional astrometry.
The second of those observations consisted of a large map
($840\arcsec\times840\arcsec$) rather than a single field of view of the camera.
\citet{mel15} performed a second map of the same size in 2015, corresponding
to the fifth epoch overall. They searched for companions to WISE 0855$-$0714
in those two maps based on common proper motions. During all of the
observations of WISE 0855$-$0714 in [4.5], images of a flanking field
were taken in [3.6]. As a result, when the two maps were obtained in [4.5],
the flanking fields in [3.6] produced maps that encompassed WISE 0855$-$0714.
To continue the astrometric monitoring, we have obtained images in [4.5]
on two additional dates in 2015.
T. Esplin et al. (in preparation) also have performed continuous imaging
of WISE 0855$-$0714 in [3.6] and [4.5] during two 23~hour periods to
characterize its variability.

We have measured astrometry for WISE 0855$-$0714 from all epochs of [4.5]
images except the two photometric monitoring campaigns.
During the monitoring for a given filter, WISE 0855$-$0714 was placed in one
corner of the array and held at a fixed location without dithering, which
optimizes photometric precision over the course of an observation but
produces larger astrometric errors than the standard observing strategy
of dithering the target near the center of the array.
To measure astrometry for WISE 0855$-$0714 from the other seven epochs, we
applied the methods described in \citet{luh14c} and the distortion corrections 
from \citet{esp16}. The resulting astrometric measurements are presented in
Table~\ref{tab:astro}.

We also have measured photometry for WISE 0855$-$0714 in [3.6] and [4.5]
from all of the data except the monitoring campaigns and the second
epoch in [3.6]. The latter set of data was excluded because WISE 0855$-$0714
was slightly blended with a brighter star. WISE 0855$-$0714 is much
brighter in [4.5], so its photometry was unaffected by the neighboring star.
For the monitoring campaigns, we have adopted the mean photometry measured
by T. Esplin et al. (in preparation). We have measured photometry from the
other data with the methods from \citet{luh12}. The resulting photometric
data are listed in Table~\ref{tab:phot}.

\subsection{VLT HAWK-I Images}
\label{sec:hawki}

We obtained images of WISE 0855$-$0714 with the High Acuity Wide-field
K-band Imager (HAWK-I) on the Unit Telescope 4 of the VLT.
The camera contains four 2048$\times$2048 HAWAII-2RG arrays and has a
plate scale of $0\farcs106$~pixel$^{-1}$ \citep{kis08}.
For these observations, we selected a medium-band filter that encompasses
the $H$-band continuum that separates two bands of CH$_4$ and H$_2$O absorption
(see Fig.~\ref{fig:filters}).
Images of WISE 0855$-$0714 in that filter were taken during portions of six
nights across a period of nearly two months (see Table~\ref{tab:log}).
The total exposure time was 6.4~hours.
Based on its proper motion and parallax (Section~\ref{sec:pm}),
WISE 0855$-$0714 moved $\sim1\arcsec$ between the first and last nights.
Therefore, when registering and combining the individual frames
from the six nights, we applied offsets to $3\arcsec\times3\arcsec$
sections of the images surrounding the expected positions of WISE 0855$-$0714
that compensated for the proper and parallactic motions.
Point sources in the final combined image exhibited FWHM$\sim0\farcs45$. 
We aligned the world coordinate system (WCS) of the image to
that of the IRAC images using offsets in right ascension, declination,
and rotation that were derived from sources detected in both sets of data.

\begin{figure}
\epsscale{1}
\plotone{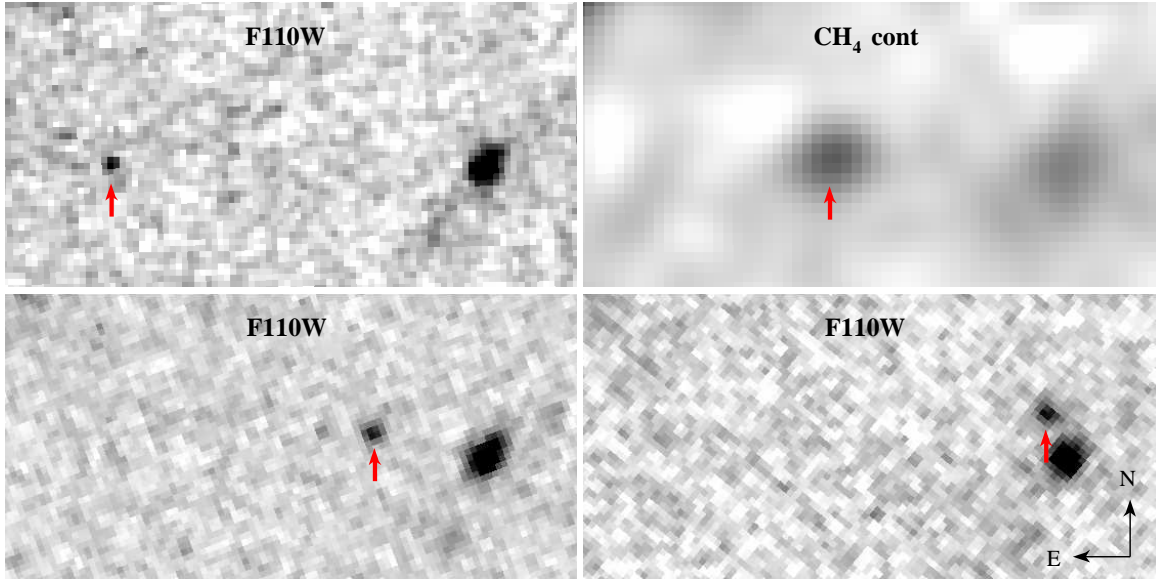}
\caption{
Images of WISE~0855$-$0714 in F110W and a CH$_4$ continuum filter
(see Figure~\ref{fig:filters}). 
The detections of WISE~0855$-$0714 are indicated by the arrows.
The F110W data were collected on three dates that spanned several months
(Table~\ref{tab:log}). The CH$_4$ continuum images also were taken
across a range of dates.  They were registered and combined
in a way that compensated for the motion of WISE~0855$-$0714.
The size of each image is $6\arcsec\times3\arcsec$.
}
\label{fig:image1}
\end{figure}

A source appears at the expected position of WISE 0855$-$0714 in the
reduced image. A section of the HAWK-I image surrounding that position
is shown in Figure~\ref{fig:image1}. The image has been smoothed to a lower
resolution of $0\farcs7$ to facilitate visual identification of the detection.
We conclude that this source is WISE 0855$-$0714 since no (stationary) object
is detected at that location in the $HST$ images, which are deeper than
the HAWK-I data.
We derived the flux calibration with $H$-band photometry from the Point Source
Catalog of the Two-Micron All-Sky Survey \citep[2MASS,][]{skr06} for sources
in the image under the assumption that they have $m_{CH4}-H\sim0$. 
Aperture photometry was then measured for WISE 0855$-$0714 using
an aperture radius of 5 pixels and radii of 7 and 11 pixels 
for the inner and outer boundaries of the sky annulus, respectively.
The photometry for WISE 0855$-$0714 is in Table~\ref{tab:phot}.
We have not used the HAWK-I image to measure astrometry for WISE 0855$-$0714
because the errors would be larger than those from the IRAC and
$HST$ images, which offer higher a signal-to-noise ratio (S/N) and higher
resolution, respectively.

\subsection{$HST$ ACS and WFC3 Images}

At the time of the planning of our initial observations with $HST$,
WISE 0855$-$0714 had been detected only at mid-IR wavelengths.
Therefore, we began by observing it in the filter on $HST$ that appeared
to offer the greatest sensitivity to cold brown dwarfs, F110W on
Wide Field Camera 3 \citep[WFC3,][]{kim08}.
The same approach was taken by \citet{luh14wd} in seeking the first
near-IR detection of the brown dwarf WD~0806-661~B.
After detecting WISE 0855$-$0714 in F110W, we pursued imaging with $HST$ in
three additional bands that are aligned with the wavelengths at
which much of the near-IR flux is predicted to emerge, consisting of
F850LP on the Advanced Camera for Surveys (ACS) and F105W and F127M on
WFC3 (see Fig.~\ref{fig:filters}).

We observed WISE 0855$-$0714 with the IR channel of WFC3 and the
Wide Field Channel (WFC) of ACS.
WFC3/IR contains a $1024\times1024$ HgCdTe array in which
the pixels have dimensions of $\sim0\farcs135\times0\farcs121$.
ACS/WFC contains two $2048\times4096$ SITe CCD arrays with plate
scales of $\sim0\farcs05$~pixel$^{-1}$.
For each of the four filters that we selected, WISE 0855$-$0714 was
observed during six orbits that were divided into three two-orbit visits.
In a given orbit, one exposure was taken at each position in a three-point
dither pattern. The dither patterns in the two orbits in
each visit were offset by 3.5 pixels along the x-axis of the array.
The position of WISE 0855$-$0714 predicted by its proper motion and parallax
was placed at the IR and WFC1-CTE apertures in WFC3 and ACS, respectively.
The WFC1-CTE aperture was selected because it is near one of the
readout amplifiers, which minimizes photometric losses due to
imperfect charge transfer efficiency. The dates and exposure times for
the visits are listed in Table~\ref{tab:log}.

\begin{figure}
\epsscale{1}
\plotone{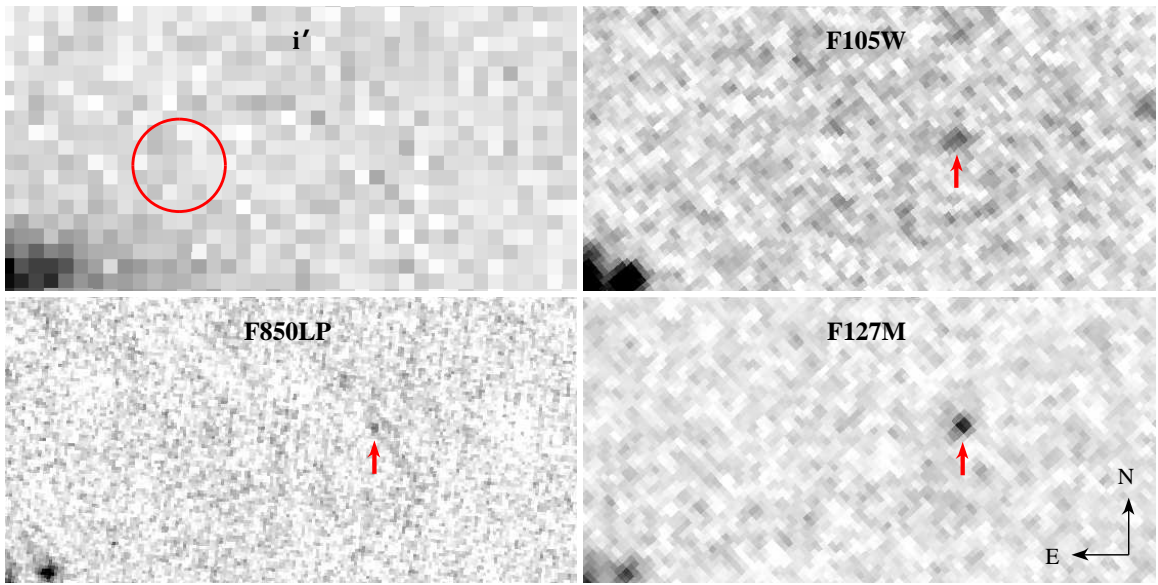}
\caption{
Images of WISE~0855$-$0714 in $i\arcmin$, F105W, F850LP, and F127M
(see Figure~\ref{fig:filters}).
The expected position of WISE~0855$-$0714 in $i\arcmin$ is circled and
the detections in the other bands are indicated by the arrows.
Each of the F850LP and F105W images is the coaddition of data from six
orbits after compensating for the object's motion. The F127M image is from
one of the three two-orbit visits in that filter.
The field covered by these images is shifted to the west
by $\sim9\arcsec$ relative to the field in Figure~\ref{fig:image1}.
The size of each image is $6\arcsec\times3\arcsec$.
}
\label{fig:image2}
\end{figure}

The WFC3 and ACS images were registered and combined using the tasks
{\it tweakreg} and {\it astrodrizzle} within the DrizzlePac software package.
We adopted drop sizes of 0.85 native pixels and resampled plate scales
of $0\farcs065$~pixel$^{-1}$ and $0\farcs035$~pixel$^{-1}$ for WFC3 and ACS,
respectively.
For F110W and F127M, we combined the six exposures within each two-orbit visit,
resulting in one reduced images for each of the three visits.
WISE 0855$-$0714 is well-detected in each of those images, as shown
in Figures~\ref{fig:image1} and \ref{fig:image2}.
Because the F127M visits spanned only one week, the movement
of WISE 0855$-$0714 among those visits was very small. As a result, 
we have included an image from only one of the three F127M visits
in Figure~\ref{fig:image2}.
Because the S/N of WISE 0855$-$0714 is low in F850LP
and F105W, we combined all exposures from the three visits for each
of those filters.  As done with the HAWK-I data, 
we corrected for the expected motion of WISE 0855$-$0714 among the visits
within a small area ($1\farcs3\arcsec\times1\farcs3$) surrounding its
expected location when registering and combining the images in F850LP
and F105W. The reduced images for those filters are shown in
Figure~\ref{fig:image2}. WISE 0855$-$0714 is clearly detected in F105W, but
only a marginal detection (S/N$\sim$3) is present in F850LP.
Because the F850LP data were taken within a few days of two of the F127M
visits, the expected position of the brown dwarf in F850LP is known precisely,
and that position does coincide with the weak source indicated in
Figure~\ref{fig:image2}.

To measure astrometry for WISE 0855$-$0714, we began by
aligning the WCS of the image from the first F110W visit to that
of the HAWK-I image, which was aligned to IRAC (Section~\ref{sec:hawki}).
We then aligned the WCS's of the other WFC3 and ACS images to the new WCS
for the first F110W visit.
The astrometry for WISE 0855$-$0714 from each of the visits in F110W
and F127M is provided in Table~\ref{tab:astro}.
We do not report astrometry from the F850LP and F105W images because
of the large errors that result from the low S/N.
As with the IRAC astrometry \citep[see][]{luh14c}, we have
estimated the astrometric errors based on the differences in
right ascension and declination between different visits for stars
with similar magnitudes as WISE 0855$-$0714.

Aperture photometry was measured for WISE 0855$-$0714 from each of the
reduced images using an aperture radius of four pixels and radii of four and
10 pixels for the inner and outer boundaries of the sky annulus, respectively.
For the WFC3 images, we measured aperture corrections of 0.097 (F105W),
0.097 (F110W), and 0.125~mag (F127M) between those apertures
and radii of $0\farcs4$ using bright stars in the images.
We then applied those corrections and the zero-point Vega magnitudes of
25.4523 (F105W), 25.8829 (F110W), and 23.4932 (F127M) for
$0\farcs4$ apertures\footnote{http://www.stsci.edu/hst/wfc3/phot\_zp\_lbn}
to the photometry of WISE 0855$-$0714. To calibrate the F850LP photometry, we
computed the zero-point STMAG magnitude for a $5\farcs5$ aperture from the
image header keyword {\it photflam} and converted it the Vega system with
the transformation from \citet{sir05}, arriving at a value of 24.316.
Because the aperture correction in F850LP is significantly larger for
redder objects, we used the software package {\it synphot}
to estimate a correction of 0.68~mag between the aperture applied
to WISE 0855$-$0714 and a $5\farcs5$ aperture.
Drizzled images can contain correlated noise, which would lead
to an underestimate of the photometric errors.
Therefore, to estimate reliable values for the errors, we
created separate versions of the reduced images that used drop sizes of
0.1 native pixels, which should minimize the correlated noise.
We then adopted the errors produced by aperture photometry on those images.
The photometry in F850LP, F105W, F110W, and F127M is presented in
Table~\ref{tab:phot}, where separate measurements are reported
for each of the three visits in F110W and F127M.

\subsection{Gemini GMOS Images}

We obtained images of WISE 0855$-$0714 in the $i\arcmin$ filter with
Gemini Multi-Object Spectrograph (GMOS) at the Gemini South telescope.
We originally proposed to conduct these observations in the $z\arcmin$ filter,
but we selected $i\arcmin$ after deeper imaging in F850LP ($\approx z\arcmin$)
with $HST$ was approved. GMOS contains three 2048$\times$4096
Hamamatsu CCD arrays that have plate scales of $0\farcs08$~pixel$^{-1}$.
WISE 0855$-$0714 was observed during portions of four nights that spanned
nearly two months with a total exposure time of 6.8~hours (see
Table~\ref{tab:log}).
The FWHM of point sources in the images ranged from 0.4--$0.8\arcsec$.
As done with other data, when registering and combining the individual
frames, we compensated for the parallactic and proper motions of
WISE 0855$-$0714 within a $3\arcsec\times3\arcsec$ section surrounding
its expected location. It was not detected in the final combined image
(see Figure~\ref{fig:image2}).
In Table~\ref{tab:phot}, we provide the magnitude limit in $i\arcmin$ that
corresponds to S/N=3.

\section{Analysis}

\subsection{Proper Motion and Parallax}
\label{sec:pm}

The proper motion and parallax of WISE 0855$-$0714 have been previously
measured by \citet{luh14b}, \citet{wri14}, and \citet{luh14c}.
For the most recent measurements, \citet{luh14c} combined 
four epochs of astrometry from {\it Spitzer} with three epochs from
{\it WISE} during its initial survey and after reactivation
\citep[{\it NEOWISE},][]{mai14} that were derived by \citet{wri14}. 
\citet{luh14c} arrived at a proper motion of 
($\mu_{\alpha}$ cos $\delta$, $\mu_{\delta}$)
=($-8.10\pm0.02$, $0.70\pm0.02\arcsec$~yr$^{-1}$)
and a parallax of $0.433\pm$0.015$\arcsec$.
We can further refine those parameters with the new
astrometry from our IRAC and $HST$ images. To do that, we employ the
two epochs of astrometry from {\it WISE} \citep{wri14} after
the adjustments by \citet{luh14c}, the astrometry that we have
measured from all seven epochs of {\it Spitzer} [4.5] images in which
WISE 0855$-$0714 was near the center of the array (i.e., excluding
the photometric monitoring campaigns), and our astrometry from
the six $HST$ visits in F127M and F110W. Those data are compiled in
Table~\ref{tab:astro}. As mentioned in Sections~\ref{sec:irac}
and \ref{sec:hawki}, we have not measured astrometry from the F850LP, F105W,
and HAWK-I images because the errors would be much larger than those for
the other data that we are utilizing.
We also exclude the {\it NEOWISE} astrometry for the same reason. 
Although the {\it WISE} data have fairly large errors, we retain them
because they significantly extend the baseline of the astrometry.

\begin{figure}
\epsscale{1.2}
\plotone{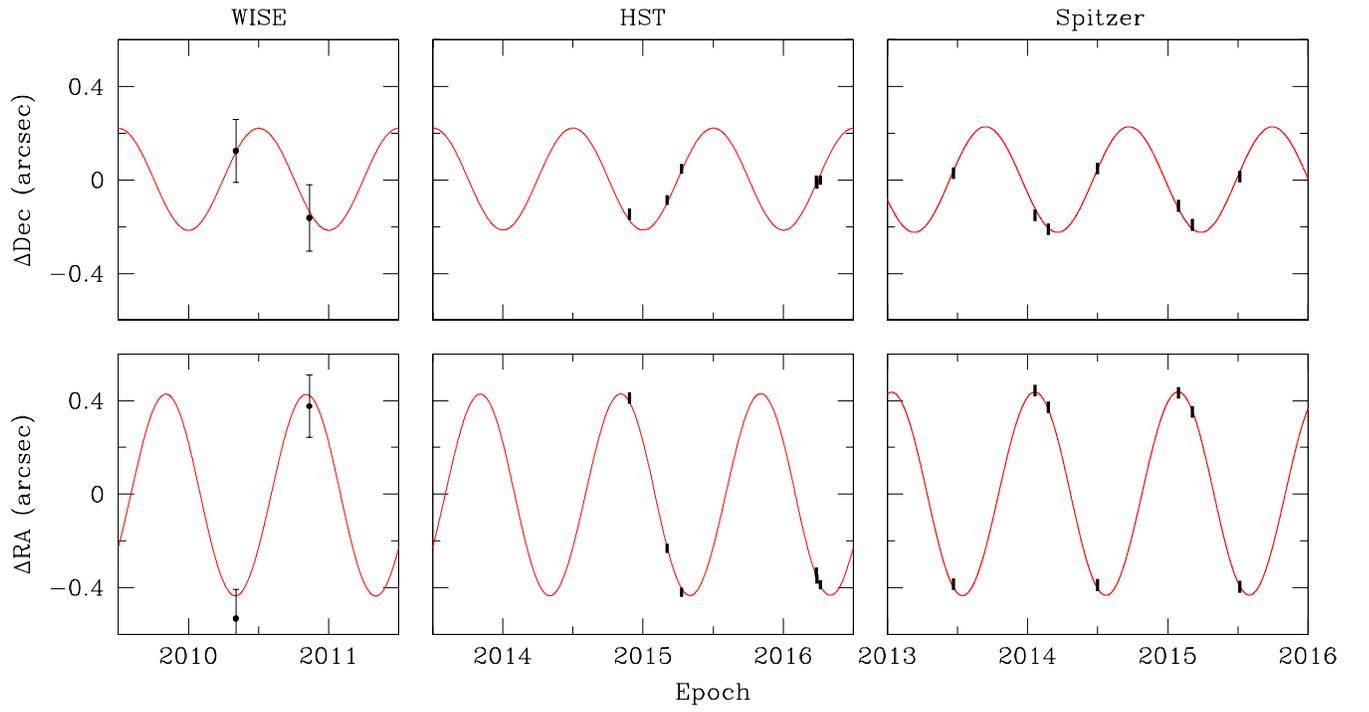}
\caption{
Relative astrometry of WISE~0855$-$0714 (Table~\ref{tab:astro}) compared
to the best-fit model of parallactic motion (red curve).
The proper motion produced by the fitting has been subtracted.
}
\label{fig:pm}
\end{figure}

We performed least-squares fitting of proper and parallactic motion to the
15 epochs of astrometry for WISE~0855$-$0714 in Table~\ref{tab:astro} with
the IDL program MPFIT. The reduced $\chi^2$ for the resulting fit was 0.4.
To verify the errors produced by the fitting, we created 10,000 sets of
astrometry that consisted of the sum of the measured astrometry and Gaussian
noise. We then fitted parallactic and proper motion to each set.
The resulting standard deviations of $\mu_{\alpha}$, $\mu_{\delta}$,
and parallax were similar to the errors from MPFIT. 
The proper motion and parallax are presented in Table~\ref{tab:prop}.
They are consistent with all of the previous estimates.
In Figure~\ref{fig:pm}, we plot the relative coordinates among the 
15 epochs after subtraction of the best-fit proper motion.

\subsection{Comparison of Observed and Model Photometry}

We wish to compare the available photometry for WISE~0855$-$0714 to
the predictions from models of the atmospheres and interiors of brown dwarfs.
A comparison of a given magnitude or color for a brown dwarf
to the model predictions can serve as a test of the validity of the
model if the defining properties of the brown dwarf are assumed,
or it can constrain the properties of the brown dwarf if the model
photometry is assumed to be accurate.
If several photometric measurements (or a spectrum) are available,
then one can pursue both a test of the models and a constraint
on the brown dwarf's properties by checking whether one suite of models
(e.g., cloudless with equilibrium chemistry)
produces a clearly superior fit, and then adopting the properties of
the best-fit model from that suite.
Such comparisons of data and models can be performed for a single object
or for a population. 
In this section, we summarize previous comparisons of photometry 
for WISE~0855$-$0714 to model predictions, compare it and
other known Y dwarfs to the models on color-magnitude diagrams (CMDs), and
perform a comparison to the full SED that we have measured for WISE~0855$-$0714.

\subsubsection{Previous Studies of WISE~0855$-$0714}
\label{sec:previous}

\citet{luh14b} compared $M_{4.5}$ and a limit on $J-4.5$ for WISE~0855$-$0714
to the predictions of multiple suites of models and identified the
temperature constraint implied by each suite for each of those measurements.
A comparison was not performed with $[3.6]-[4.5]$ because theoretical values
differ greatly from the observed colors of T and Y dwarfs \citep{leg10,bei14}.
The models considered by \citet{luh14b} were defined primarily by the
following features: water clouds and chemical equilibrium \citep{bur03},
cloudless and chemical equilibrium \citep{sau08,sau12},
cloudless and non-equilibrium chemistry \citep{sau08,sau12}, and
50\% coverage of water, chloride, and sulfide clouds and chemical equilibrium
\citep{mor12,mor14}. In our study, we employ the same models except for
those from \citet{bur03}, which are omitted because of their poorer agreement
with the data for WISE~0855$-$0714 and other Y dwarfs \citep{liu12,luh14wd}.

\citet{fah14} obtained images of WISE~0855$-$0714 in a medium-band filter
within the $J$ band, achieving a possible 2.6~$\sigma$ detection.
They used their photometry to estimate the $J$ magnitude of WISE~0855$-$0714
and place it on a diagram of $M_{W2}$ versus $J-W2$, where $W2$ is a {\it WISE}
band that is similar to [4.5] from {\it Spitzer}. 
In that diagram, the brown dwarf appeared closer to the cloudy models
from \citet{mor12,mor14} than the cloudless/chemical equilibrium models
from \citet{sau12}, which they cited as evidence of water ice clouds.
However, using the $J$ estimate from that study, \citet{luh14c} found that 
WISE~0855$-$0714 was roughly midway between the cloudless and cloudy models
that assume chemical equilibrium in a diagram of $M_{4.5}$ versus $J-[4.5]$,
and that it agreed with cloudless models that invoke non-equilibrium chemistry
\citep{sau08,sau12}.
In general, to draw a definitive conclusion regarding the physical properties
of a brown dwarf (e.g., presence of clouds) from a single CMD,
one must assume that each suite of models produces colors that are accurate
for the physical conditions in question.
Such an assumption is not warranted given the
untested nature of those colors near the temperature of WISE~0855$-$0714.
Indeed, the suite of models that best matches the data for WISE~0855$-$0714
and other Y dwarfs varies among different CMDs, as shown in the next section.

\citet{bea14} measured a limit on the $Y$-band magnitude of WISE~0855$-$0714.
They compared the updated SED to their new version of the BT-Settl models and
the cloudy models from \citet{mor14}. The best fit models had
temperatures of 240 and 250~K, respectively, both with log~$g=4$.
The $Y$-band limit from \citet{bea14} is similar to the brightest fluxes
predicted at the observed value of $M_{4.5}$ for WISE~0855$-$0714
from among the suites of models that we consider.

\citet{kop14} obtained $z$-band images of WISE~0855$-$0714 that did not
show a detection. They compared the SED that included their limit
to models of cloudy brown dwarfs from \citet{bur03} and \citet{mor14}
for a range of temperatures and surface gravities. During the comparison,
each model SED was scaled to match the observed fluxes in [3.6] and [4.5],
which was equivalent to scaling the radius.
However, that approach can lead to radii that differ significantly from
those predicted by evolutionary models for a given temperature and surface
gravity \citep{bur03,sau08}, and thus are unphysical. 
Indeed, among the eight best-fit models that they presented,
\citet{kop14} noted that five models exhibited unphysical values of
radius (0.4--0.5~$M_{\rm Jup}$) or surface gravity (log~$g>4.5$).
In fact, the gravity and radius of one of their models (log~$g=5.5$,
1.4~$M_{\rm Jup}$) corresponded to a stellar mass (0.23~$M_\odot$). 
We find that the best-fit radii of two of their three remaining models
also differ from those predicted by evolutionary models by 30--50\%.
Based on their comparison to the theoretical SEDs, \citet{kop14} concluded
that no models reproduced the $[3.6]-[4.5]$ color of WISE~0855$-$0714, which
was previously known to apply to T and Y dwarfs in general \citep{leg10,bei14}.

\subsubsection{Color-Magnitude Diagrams}

We can use CMDs to place the photometry of WISE~0855$-$0714 in the
context of data for other brown dwarfs and to compare trends within this
population to model predictions. Because few data are available for Y dwarfs in
optical bands \citep{lod13,kop14,leg15}, we have not constructed CMDs with
$i\arcmin$ and F850LP ($\approx z\arcmin$).
Previous data for F105W, F110W, F127M, and the CH$_4$ continuum filter
are also limited, but most known Y dwarfs have been observed in the
overlapping bands of $Y$, $J$, and $H$ (see Fig.~\ref{fig:filters}).
Therefore, we select the latter three bands and the two {\it Spitzer} filters
for our CMDs. We plot $M_{4.5}$ on the vertical axis of each diagram because
it encompasses less absorption and exhibits higher S/N for Y dwarfs
than the other filters, and because it captures most of the flux of
Y dwarfs at $<5$~\micron. 

\begin{figure}
\epsscale{1}
\plotone{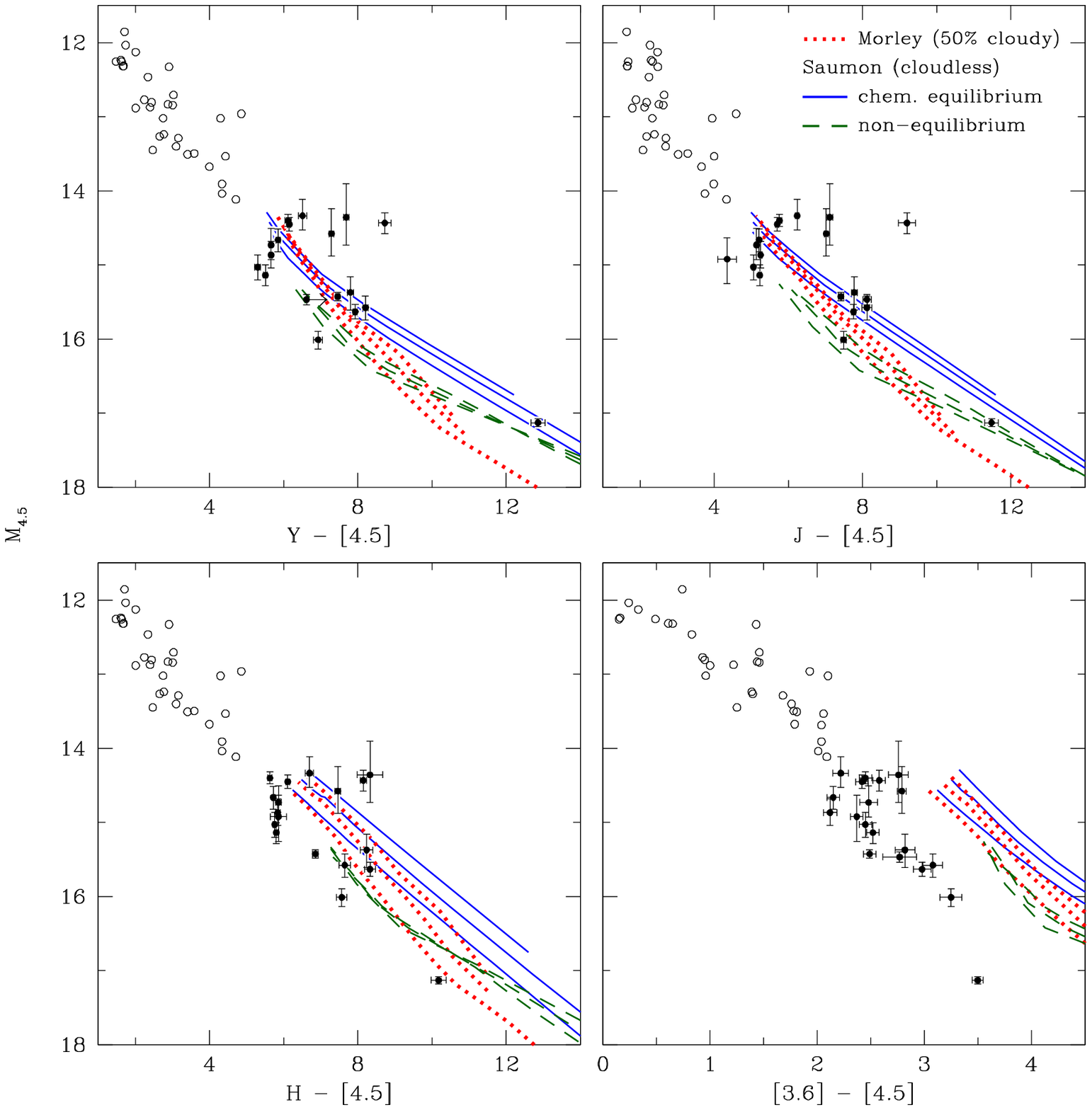}
\caption{
Color-magnitude diagrams for WISE~0855$-$0714 (faintest point)
and samples of T dwarfs \citep[open circles,][references therein]{dup12}
and Y dwarfs \citep[filled circles with error
bars,][]{cus11,cus14,tin12,tin14,luh12,luh14wd,bei13,bei14,kir12,kir13,dup13,leg13,leg15,leg16,sch15}.
These data are compared to the magnitudes and colors predicted
by three sets of theoretical models (red solid, blue dotted, and
green dashed lines) for ages of 1, 3, and 10~Gyr \citep{sau12,mor12,mor14}.
The $Y$, $J$, and $H$ magnitudes for WISE~0855$-$0714 have been estimated
by combining our measurements in F105W, F127M, and the CH$_4$ continuum
filter with the values of $Y-m_{105}$, $J-m_{127}$, and $H-m_{CH4}$
predicted by these models.
}
\label{fig:cmd1}
\end{figure}

\begin{figure}
\epsscale{1}
\plotone{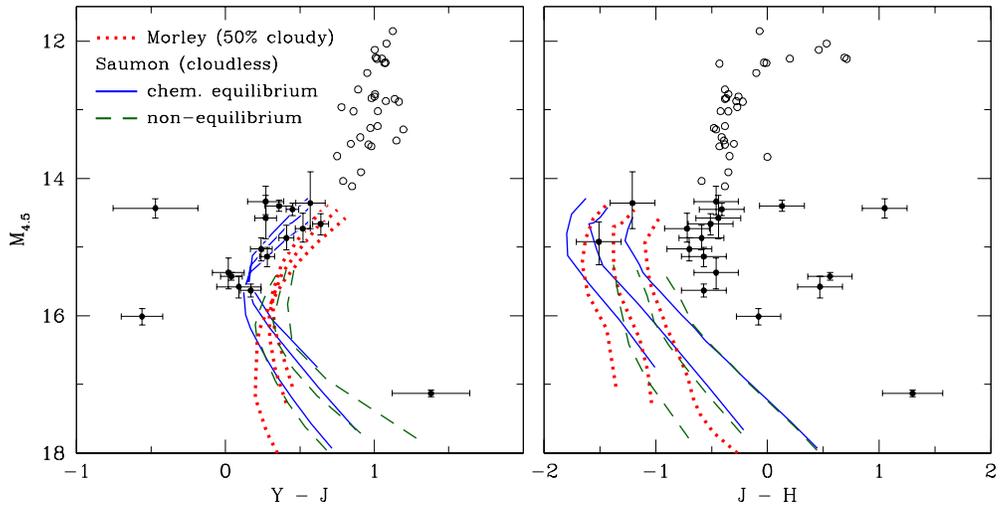}
\caption{
Same as Figure~\ref{fig:cmd1} for $Y-J$ and $J-H$.
}
\label{fig:cmd2}
\end{figure}

In Figure~\ref{fig:cmd1}, we show CMDs
with colors that extend from $Y/J/H/[3.6]$ to [4.5].
We also plot CMDs with $Y-J$ and $J-H$ in Figure~\ref{fig:cmd2}.
To place WISE~0855$-$0714 in these CMDs, we have combined our
measurements of $m_{105}$, $m_{127}$, and $m_{CH4}$ with 
$Y-m_{105}$, $J-m_{127}$, and $H-m_{CH4}$ as predicted by the models
at the value of $M_{4.5}$ for WISE~0855$-$0714. The three suites of models that
we consider (Section~\ref{sec:previous}) produce similar values of $J-m_{127}$
($\sim$0.89) and $H-m_{CH4}$ ($\sim$0.85), but the predicted $Y-m_{105}$ ranges 
from $-$0.25 to $-$0.9. We adopted $Y-m_{105}=-0.6$ for plotting 
WISE~0855$-$0714 in Figure~\ref{fig:cmd1}. Its position relative to the models
in the $Y-[4.5]$ CMD does not change significantly if a different value
is adopted given that the model suites span more than 3~mag in $Y-[4.5]$
at the magnitude of WISE~0855$-$0714.
For $m_{127}$ and [4.5], we have adopted the means of the multiple measurements
that are available. WISE~0855$-$0714 is plotted with an error bar in
$J-[4.5]$ that corresponds to the range of the three $m_{127}$ measurements.
For $[3.6]-[4.5]$, we have adopted the mean color 
that was measured in the four epochs with [3.6] data. 
Because model colors have been used to convert from
$m_{105}/m_{127}/m_{CH4}$ to $Y/J/H$ for WISE~0855$-$0714, the CMDs
are technically showing the positions of WISE~0855$-$0714 relative
to the models in $m_{105}/m_{127}/m_{CH4}-[4.5]$ and the positions
of other Y dwarfs relative to the models in $Y/J/H-[4.5]$.

In the CMDs, we have plotted a sample of T dwarfs
\citep[][references therein]{dup12} and all known Y dwarfs that have
measurements of parallaxes and photometry in relevant bands
\citep{cus11,cus14,tin12,tin14,luh12,luh14wd,bei13,bei14,kir12,kir13,dup13,leg13,leg15,leg16,sch15}.
For the Y dwarf WISE J035000.32$-$565830.2 (hereafter WISE 0350$-$5658),
the parallax of $0.291\pm0.50\arcsec$ from \citet{mar13} places it at a
discrepant location in CMDs relative to other Y dwarfs \citep{leg15,leg16}.
Meanwhile, four known Y dwarfs lack previous parallax measurements,
consisting of WISE J030449.03-270508.3 \citep{pin14},
WISE J082507.35+280548.5 (hereafter WISE 0825+2805), WISE J120604.38+840110.6
(hereafter WISE 1206+8401), and WISE J235402.77+024015.0 \citep{sch15}.
A significant amount of multi-epoch {\it Spitzer} data is now publicly available
for WISE 0350$-$5658, WISE 0825+2805, and WISE 1206+8401.
To resolve the discrepancy in the CMD locations of WISE 0350$-$5658
and to allow the addition of WISE 0825+2805 and WISE 1206+8401 to the CMDs,
we have measured their parallaxes from the {\it Spitzer} data
with the same methods that were applied to WISE~0855$-$0714, arriving
at $0.184\pm0.01$, $0.158\pm0.007$, and $0.085\pm0.007\arcsec$, respectively.
It is likely that these values will be soon superseded by measurements
that include the {\it Spitzer} images that are not yet publicly available
as well as astrometry from other telescopes. Using our new parallaxes,
all three of these objects exhibit locations in the CMDs that are
consistent with the sequences formed by other Y dwarfs.

For comparison to the data in the CMDs, we have included 
magnitudes and colors predicted by three suites of models that were
mentioned in Section~\ref{sec:previous}: cloudless with chemical equilibrium,
cloudless with non-equilibrium chemistry, and 50\% cloud coverage
with chemical equilibrium  \citep{sau08,sau12,mor12,mor14}.
The ages of the Y dwarfs in the CMDs are unknown with the exception of
WD~0806-661~B \citep[2$\pm$0.5~Gyr,][]{luh12}, so we show model isochrones for
ages of 1, 3, and 10~Gyr, which span the ages of most stars in the solar
neighborhood. The models with equilibrium and non-equilibrium chemistry
are plotted for temperatures of $<450$~K and $<350$~K, respectively.

In Figure~\ref{fig:cmd1}, the sequences formed by known Y dwarfs are
qualitatively similar among the CMDs with $Y-[4.5]$, $J-[4.5]$, and $H-[4.5]$.
The CMD with $[3.6]-[4.5]$ contains the most well-defined sequence
because [3.6] and [4.5] usually offer the most accurate photometry for Y dwarfs
from available bands.
Because of their large scatter, the sequences of Y dwarfs
brighter than WISE~0855$-$0714 in $Y/J/H-[4.5]$ do not tightly constrain
the models, but those sequences are in rough agreement with all three model
suites, agreeing somewhat better with the cloudless/chemical equilibrium
models in $Y-[4.5]$ and $J-[4.5]$. All of the models are much redder
than the data in $[3.6]-[4.5]$, as found in a number of previous studies
\citep{leg10,bei14,luh14wd}.
It is unfortunate that the CMD with the best-defined Y dwarf sequence contains
a color that is especially difficult for the models to reproduce.
The differences between the three suites of models in the CMDs
of $Y/J/H-[4.5]$ increase at fainter magnitudes, so 
WISE~0855$-$0714 offers the greatest potential for discriminating among those
models. However, the suite of models that best matches the position of
WISE~0855$-$0714 changes from one CMD to the next:
cloudless/chemical equilibrium in $Y-[4.5]$, cloudless/non-equilibrium
chemistry in $J-[4.5]$, and cloudy in $H-[4.5]$. For the $J-[4.5]$ CMD,
\citet{luh14c} arrived at a similar result using the $J$-band measurement
from \citet{fah14}.
We note that the overall agreement between the Y dwarf sequences
and the model isochrones is not improved by replacing [4.5] with a
different band on the vertical axes in the CMDs.

The $Y-J$ CMD in Figure~\ref{fig:cmd2} exhibits a well-defined sequence
of Y dwarfs plus two blue outliers, WISE 0350$-$5658 and
WISE 182831.08+265037.8. The sequence
of late-T and Y dwarfs becomes bluer at fainter magnitudes, as found in 
many previous studies \citep{liu12,dup13,mor14,sch15,leg13,leg15,leg16}.
That trend has been attributed to the depletion of neutral alkalis into
gases and solids \citep{liu12}, which would reduce the absorption from
pressure-broadened alkali lines at red optical wavelengths \citep{mar02,bur03}.
The cloudless models with chemical equilibrium reproduce the Y dwarf
sequence in $Y-J$ while the isochrones from the other two suites of models
are too red \citep{liu12,leg13,leg15,mor14}.
\citet{tre15} found that their cloudless models produced similar $Y-J$ colors
for equilibrium and non-equilibrium chemistry, both of which matched the data
for Y dwarfs.
Whereas Y dwarfs have previously exhibited bluer $Y-J$ at fainter magnitudes,
our data imply a redder color for WISE~0855$-$0714.
A similar result is found when WISE~0855$-$0714 is compared to other Y dwarfs
in terms of $m_{105}-J$ using F105W photometry from \citet{sch15}.
As shown in Figure~\ref{fig:cmd2}, all three sets of models do predict
a shift of the Y dwarf sequence back to redder values of $Y-J$ for objects
as faint as WISE~0855$-$0714.

As in the $Y-J$ CMD, a clear extension of the T dwarf sequence is evident
among the Y dwarfs in the $J-H$ CMD, except with more discrepant objects
in the latter. The blue outliers are WISE J053516.80$-$750024 and
WISE J014656.66+423410.0 and the reddest outlier is WISE 182831.08+265037.8.
Previous studies have compared similar $J-H$ CMDs (often with $M_J$)
to model predictions \citep{mor12,mor14,leg13,leg15,leg16,mar13,bei14},
finding that the cloudless and partly cloudy models of \citet{sau12}
and \citet{mor12,mor14} tend to produce colors that are too blue
(see also Figure~\ref{fig:cmd2}). Those models agree with the observed
colors only if the surface gravity is increased to a value that becomes
unphysical for cooler Y dwarfs (log~$g\gtrsim5$) and the cloud coverage
is very large \citep{mor14}. The cloudless non-equilibrium models in
Figure~\ref{fig:cmd2} are also too blue, although those from \citet{tre15}
are red enough to match the $J-H$ data \citep{leg16}.
Previous studies have shown that Y dwarfs start to become redder in $J-H$
at fainter magnitudes \citep{sch15,leg15,leg16}, 
and we find that WISE~0855$-$0714 continues that trend, which agrees with
the model predictions (see Figure~\ref{fig:cmd2}).

\subsubsection{Spectral Energy Distribution}

In Figure~\ref{fig:dm}, we have constructed an SED for WISE~0855$-$0714 from
our photometric measurements in $i\arcmin$, F850LP, F105W, F110W, F127M,
the CH$_4$ continuum filter, [3.6], and [4.5]. The other available data
in Table~\ref{tab:phot} are omitted from the SED because they are
superseded by our more sensitive photometry in similar bands or they
are not deep enough to provide a useful constraint on the models ($K_s$).
As in Figure~\ref{fig:filters}, we include in Figure~\ref{fig:dm} an example of
a model spectrum of a cold brown dwarf \citep{mor14} to indicate the
wavelengths of major spectral features relative to the bands of our photometry.

For each of the three model suites considered in this work and for each of the
ages of 1, 3, and 10~Gyr, we have identified the model that has the
same value of $M_{4.5}$ as WISE~0855$-$0714 because this band 
encompasses most of the flux at $<$5~\micron.
For each of those nine models, we then computed the difference between the
observed and predicted absolute magnitudes for each of the bands in the
SED for WISE~0855$-$0714.
In Figure~\ref{fig:dm}, we plot the resulting differences for the three
model suites at 3~Gyr. The temperatures for those best-fit models are
237, 244, an 249~K for cloudless/chemical equilibrium, cloudless/non-equilibrium
chemistry, and 50\% cloudy, respectively. The other two ages have been omitted
because, for a given model suite, most of the magnitude differences for
1, 3, and 10~Gyr span a modest range on the scale of Figure~\ref{fig:dm}
($\lesssim0.5$~mag).
Positive and negative deviations correspond to predicted fluxes that are
brighter and fainter than the data, respectively.

For all of the model suites considered, the predicted SEDs exhibit large
deviations from the observed SED of WISE~0855$-$0714.
No single model provides a clearly superior match to the data.
When the models are selected to match the object at [4.5] as we have done,
they are too faint in $H$ and [3.6]. That is a reflection of the fact that
the models are too red in $H-[4.5]$ and $[3.6]-[4.5]$, as found in the CMDs.
The deviation in $H$ is smallest for the cloudy models, which
is also evident from the $H-[4.5]$ CMD in Figure~\ref{fig:cmd1}.
However, the two suites of cloudless models agree better with the
SED of WISE~0855$-$0714 in F850LP through F127M. The cloudy models
are too bright by 1.5--2.5~mag in those bands, where the deviation becomes
progressively larger at shorter wavelengths.

If one selected a model that matched WISE~0855$-$0714 in a different band
than [4.5], all of the deviations in Figure~\ref{fig:dm} would shift vertically
in the same direction so that the new normalization band exhibited zero
deviation. In addition, because the temperature of that model would
differ from that derived by fitting to [4.5] and because the near-IR fluxes
are more sensitive to temperature than the flux in [4.5], the
shifts of the deviations would be larger in the near-IR bands.
As with the CMDs, adopting a different band than [4.5] in this analysis does
not improve the agreement between the observed SED and the model predictions.

\subsection{Multiplicity Constraints}

Among the images in which WISE~0855$-$0714 has been detected, those
from $HST$ offer the highest resolution, and therefore can detect a companion
at the smallest separations. The F110W and F127M images
have provided the highest S/N for WISE~0855$-$0714 from among the four filters
with which it was observed by $HST$. The S/N in each of those filters is
similar to that of the brown dwarf WD~0806-661~B in F110W images from
\citet{luh14wd}. Thus, the constraints on the presence of a companion
to WISE~0855$-$0714 are similar to those derived from the images
of WD~0806-661~B, which were capable of detecting $\sim$80\% of companions
with $\Delta m_{110}\lesssim0.7$~mag at separations beyond $0\farcs13$
($\gtrsim0.3$~AU for WISE~0855$-$0714).

Because of improvements in the accuracies of parallaxes for Y dwarfs
\citep[][Section~\ref{sec:pm}]{dup13,bei14,tin14}, the Y dwarf sequence
in the $[3.6]-[4.5]$ CMD (Figure~\ref{fig:cmd1}) is now sufficiently narrow
and well-defined that one could attempt to identify unresolved binaries
via their elevated positions relative to the sequence. That method is not
yet applicable to WISE~0855$-$0714 since no other Y dwarfs have been
found near its absolute magnitude.

\section{Discussion}

Because of its low temperature, WISE~0855$-$0714 is
very red from near- to mid-IR wavelengths.
Its previous detections were limited to one near-IR band (S/N=2.6)
and two mid-IR bands.
We have presented deep images of WISE~0855$-$0714 in six optical and
near-IR filters, five of which show detections (one is marginal at S/N$\sim3$).
We also have continued our previous mid-IR imaging of WISE~0855$-$0714 to
refine its parallax.

Using our new photometry and parallax measurement, we have placed
WISE~0855$-$0714 in several CMDs and constructed its SED. 
For comparison, we have included in the CMDs other known Y dwarfs
and the isochrones predicted by three suites of models for the coldest
brown dwarfs \citep{sau12,mor12,mor14}.
Previous studies have found that Y dwarfs at fainter magnitudes
become bluer in $Y-J$ and begin to become redder in $J-H$ 
\citep[e.g.,][]{liu12,sch15,leg16}.
WISE~0855$-$0714 is redder than other Y dwarfs in both colors, which
confirms the prediction that the near-IR colors of brown dwarfs turn
to redder values at temperatures below $\sim$300--400~K as the 
Wien tail increasingly affects the colors \citep{bur03,sau12,mor14}.
No single model is able to reproduce the sequences formed by 
WISE~0855$-$0714 and other Y dwarfs in all CMDs.
Similarly, the SEDs predicted by all three model suites differ significantly
from the SED of WISE~0855$-$0714.
When normalized to the latter in the [4.5] band, the model SEDs are too
faint in $H$ and [3.6], which has been found previously for other brown
dwarfs \citep{leg10,mor14}. The deviations at shorter wavelengths exhibit
a trend with wavelength, and are particularly large for the cloudy models,
which are 1.5--2.5~mag too bright.
Because none of the models provide a good match to the
SED of WISE~0855$-$0714, we are unable to determine whether clouds or
non-equilibrium chemistry are likely to be present in WISE~0855$-$0714.

\begin{figure}
\epsscale{1}
\plotone{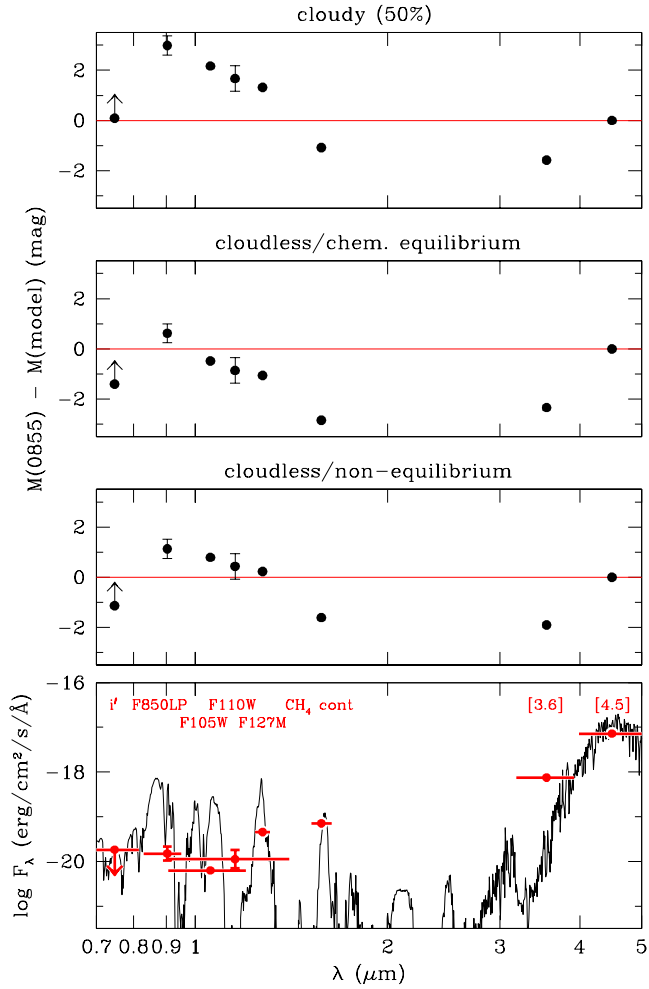}
\caption{
Bottom: SED of WISE~0855$-$0714 (red points) and an example of a theoretical
spectrum of a brown dwarf at the distance and $M_{4.5}$ of WISE~0855$-$0714
\citep[50\% cloud coverage, 250~K,][]{mor14}. 
The horizontal bars on the data points represent the widths of the filters.
The vertical bars represent the error in F850LP and the range of the
multiple measurements in F110W. The errors in the other bands are 
smaller than the data points.
Top three panels: Differences in absolute magnitudes between WISE~0855$-$0714
and models that match its $M_{4.5}$ for an age of 3~Gyr
\citep{sau12,mor12,mor14}. A point above/below zero indicates that the
predicted flux is brighter/fainter than the observed value.
}
\label{fig:dm}
\end{figure}

It is not surprising that the predicted and observed SEDs of WISE~0855$-$0714
differ significantly. The models have not been previously
tested at such low temperatures, and several aspects of the models are
uncertain, which include the methane opacities, the molecules that are
affected by non-equilibrium chemistry, and various aspects regarding the
treatment of water clouds \citep{bur03,sau12,mor14}.
The opacities from the pressure-broadened alkali lines are also uncertain,
although those lines should be weak at the temperature of WISE~0855$-$0714
because of depletion of the alkalis into condensates \citep{bur03,mor14}.
The wavelength dependence of the deviations of the models from the data
at $<1.5$~\micron\ may provide a clue to their origin (see Figure~\ref{fig:dm}).
The current measurement of the SED for WISE~0855$-$0714 should serve as
a valuable test of newer generations of models.

We conclude by discussing the prospects for future observations of
WISE~0855$-$0714.
We have presented photometry that covers most of the wavelength range from
0.7--5~\micron, so imaging in additional near-IR bands is not essential.
Our multiple epochs of photometry in F110W and F127M may show evidence
of variability, which could be used to detect clouds and hot spots
if measured in greater detail \citep{mor14b}. However, it may not be
feasible to measure time-series photometry with sufficient accuracy and
cadence given that a single $HST$ detection with S/N$\sim$10 requires
a few hours of observing time.
Near-IR spectroscopy of WISE~0855$-$0714 may be possible with $HST$.
Based on our photometry and previous spectroscopy of Y dwarfs
\citep[e.g.,][]{cus11,sch15}, a minimum of 20--30~orbits would be needed
to reach S/N$\sim5$ in a low-resolution near-IR spectrum of WISE~0855$-$0714.
The {\it James Webb Space Telescope} will be capable of obtaining near-
and mid-IR spectra of it with significantly higher S/N and resolution
\citep{mor14}.

\acknowledgements
We thank Caroline Morley and Didier Saumon for providing their model
calculations.
We acknowledge support from grants GO-13802 and GO-14157 from the Space
Telescope Science Institute and a grant from NASA issued by the 
Jet Propulsion Laboratory (JPL), California Institute of Technology.
The {\it Spitzer Space Telescope} is operated by JPL/Caltech
under a contract with NASA.
The NASA/ESA {\it Hubble Space Telescope} is operated by the Space Telescope
Science Institute and the Association of Universities for Research in
Astronomy, Inc., under NASA contract NAS 5-26555.
Gemini Observatory is operated by AURA under a cooperative agreement with
the NSF on behalf of the Gemini partnership: the NSF (United States), the NRC
(Canada), CONICYT (Chile), the ARC (Australia),
Minist\'{e}rio da Ci\^{e}ncia, Tecnologia e Inova\c{c}\~{a}o (Brazil) and
Ministerio de Ciencia, Tecnolog\'{i}a e Innovaci\'{o}n Productiva (Argentina).
2MASS is a joint project of the University of Massachusetts and the Infrared
Processing and Analysis Center (IPAC) at Caltech, funded by NASA and the NSF.
The Center for Exoplanets and Habitable Worlds is supported by the
Pennsylvania State University, the Eberly College of Science, and the
Pennsylvania Space Grant Consortium.

\clearpage

\begin{deluxetable}{lllll}
\tabletypesize{\scriptsize}
\tablewidth{0pt}
\tablecaption{Observing Log\label{tab:log}}
\tablehead{
\colhead{Telescope/Instrument} & \colhead{Filter} & \colhead{N$\times\tau_{\rm exp}$} & \colhead{Date\tablenotemark{a}} & \colhead{Program ID}\\
\colhead{} & \colhead{} & \colhead{(sec)} & \colhead{} & \colhead{}}
\startdata
{\it Spitzer}/IRAC & [3.6] & 5$\times$23.6 & 2013 Jun 21 & 90095 \\
{\it Spitzer}/IRAC & [4.5] & 5$\times$26.8 & 2013 Jun 21 & 90095 \\
{\it Spitzer}/IRAC & [4.5] & 5$\times$26.8 & 2014 Jan 20 & 90095 \\
{\it Spitzer}/IRAC & [4.5] & 5$\times$26.8 & 2014 Feb 24 & 90095 \\
{\it Spitzer}/IRAC & [3.6] & 9$\times$26.8 & 2014 Jul 1 & 10168 \\
{\it Spitzer}/IRAC & [4.5] & 9$\times$26.8 & 2014 Jul 1 & 10168 \\
{\it HST}/WFC3 & F110W & 6$\times$903 & 2014 Nov 25 & 13802 \\
VLT/HAWK-I & CH$_4$ cont & 50$\times$100 & 2014 Dec 1 & 094.C-0048 \\
VLT/HAWK-I & CH$_4$ cont & 25$\times$100 & 2015 Jan 9 & 094.C-0048 \\
VLT/HAWK-I & CH$_4$ cont & 25$\times$100 & 2015 Jan 15 & 094.C-0048 \\
VLT/HAWK-I & CH$_4$ cont & 50$\times$100 & 2015 Jan 17 & 094.C-0048 \\
VLT/HAWK-I & CH$_4$ cont & 50$\times$100 & 2015 Jan 19 & 094.C-0048 \\
VLT/HAWK-I & CH$_4$ cont & 30$\times$100 & 2015 Jan 20 & 094.C-0048 \\
{\it Spitzer}/IRAC & [3.6] & 9$\times$26.8 & 2015 Jan 29 & 10168 \\
{\it Spitzer}/IRAC & [4.5] & 9$\times$26.8 & 2015 Jan 29 & 10168 \\
{\it HST}/WFC3 & F110W & 6$\times$903 & 2015 Mar 3 & 13802 \\
{\it Spitzer}/IRAC & [4.5] & 5$\times$26.8 & 2015 Mar 5 & 10168 \\
{\it Spitzer}/IRAC & [3.6] & 405$\times$93.6 & 2015 Mar 10 & 11056 \\
{\it Spitzer}/IRAC & [4.5] & 405$\times$96.8 & 2015 Mar 9 & 11056 \\
{\it HST}/WFC3 & F110W & 6$\times$903 & 2015 Apr 11 & 13802 \\
{\it Spitzer}/IRAC & [4.5] & 5$\times$26.8 & 2015 Jul 6 & 10168 \\
{\it Spitzer}/IRAC & [3.6] & 405$\times$93.6 & 2015 Aug 3 & 11056 \\
{\it Spitzer}/IRAC & [4.5] & 405$\times$96.8 & 2015 Aug 3 & 11056 \\
Gemini South/GMOS & $i\arcmin$ & 12$\times$609 & 2015 Nov 21 & GS-2015B-Q-16 \\
Gemini South/GMOS & $i\arcmin$ & 4$\times$609 & 2015 Dec 13 & GS-2015B-Q-16 \\
Gemini South/GMOS & $i\arcmin$ & 7$\times$609 & 2016 Jan 9 & GS-2015B-Q-16 \\
Gemini South/GMOS & $i\arcmin$ & 17$\times$609 & 2016 Jan 10 & GS-2015B-Q-16 \\
{\it HST}/WFC3 & F105W & 6$\times$903 & 2016 Mar 22 & 14157 \\
{\it HST}/ACS & F850LP & 6$\times$810 & 2016 Mar 23 & 14157 \\
{\it HST}/WFC3 & F105W & 6$\times$903 & 2016 Mar 23 & 14157 \\
{\it HST}/ACS & F850LP & 6$\times$810 & 2016 Mar 24 & 14157 \\
{\it HST}/ACS & F850LP & 6$\times$810 & 2016 Mar 24 & 14157 \\
{\it HST}/WFC3 & F127M & 6$\times$903 & 2016 Mar 27 & 14157 \\
{\it HST}/WFC3 & F105W & 6$\times$903 & 2016 Mar 28 & 14157 \\
{\it HST}/WFC3 & F127M & 6$\times$903 & 2016 Mar 28 & 14157 \\
{\it HST}/WFC3 & F127M & 6$\times$903 & 2016 Apr 6 & 14157 
\enddata
\tablenotetext{a}{For VLT and Gemini, the dates apply to the
beginning of the nights of the observations. For {\it Hubble} and
{\it Spitzer}, the UT dates of the observations are listed.}
\end{deluxetable}

\begin{deluxetable}{llllll}
\tabletypesize{\scriptsize}
\tablewidth{0pt}
\tablecaption{Astrometry of WISE~J085510.83$-$071442.5\label{tab:astro}}
\tablehead{
\colhead{$\alpha$ (J2000)} & \colhead{$\sigma_{\alpha}$} & \colhead{$\delta$ (J2000)} & \colhead{$\sigma_{\delta}$} & \colhead{MJD} &  \colhead{Source} \\
\colhead{($\arcdeg$)} & \colhead{($\arcsec$)} & \colhead{($\arcdeg$)} & \colhead{($\arcsec$)} & \colhead{} & \colhead{}}
\startdata
133.7952573 & 0.125 & $-$7.2450910 & 0.135 & 55320.38 & {\it WISE} \\
133.7943232 & 0.133 & $-$7.2450719 & 0.142 & 55511.35 & {\it WISE} \\
133.7881778 & 0.025 & $-$7.2445255 & 0.025 & 56464.51 & {\it Spitzer} \\
133.7870855 & 0.025 & $-$7.2444659 & 0.025 & 56677.29 & {\it Spitzer} \\
133.7868470 & 0.025 & $-$7.2444642 & 0.025 & 56712.35 & {\it Spitzer} \\
133.7858416 & 0.025 & $-$7.2443262 & 0.025 & 56839.71 & {\it Spitzer} \\
133.7851503 & 0.025 & $-$7.2443041 & 0.025 & 56986.82 & {\it HST} \\
133.7847553 & 0.025 & $-$7.2442609 & 0.025 & 57051.25 & {\it Spitzer} \\
133.7843602 & 0.020 & $-$7.2442366 & 0.020 & 57084.82 & {\it HST} \\
133.7845153 & 0.025 & $-$7.2442655 & 0.025 & 57086.16 & {\it Spitzer} \\
133.7840688 & 0.020 & $-$7.2441796 & 0.020 & 57123.17 & {\it HST} \\
133.7835385 & 0.025 & $-$7.2441445 & 0.025 & 57209.43 & {\it Spitzer} \\
133.7819058 & 0.020 & $-$7.2440115 & 0.020 & 57474.62 & {\it HST} \\
133.7818923 & 0.020 & $-$7.2440157 & 0.020 & 57475.41 & {\it HST} \\
133.7818300 & 0.020 & $-$7.2440063 & 0.020 & 57484.38 & {\it HST} 
\enddata
\tablecomments{The {\it WISE} data are from \citet{wri14}
after the adjustments described by \citet{luh14c}.}
\end{deluxetable}

\begin{deluxetable}{lll}
\tabletypesize{\scriptsize}
\tablewidth{0pt}
\tablecaption{Photometry of WISE~J085510.83$-$071442.5\label{tab:phot}}
\tablehead{
\colhead{Band} & \colhead{Magnitude} & \colhead{Reference}}
\startdata
$i\arcmin$ & $>$27.2\tablenotemark{a} & 1 \\
$z\arcmin$ & $>$24.3\tablenotemark{a} & 2 \\
$m_{850}$ & 26.85$^{+0.31}_{-0.44}$ & 1 \\
$Y$ & $>$24.4\tablenotemark{a} & 3 \\
$m_{105}$ & 27.33$\pm$0.19 & 1 \\
$m_{110}$ & 26.71$\pm$0.19 & 1 \\
$m_{110}$ & 26.47$\pm$0.13 & 1 \\
$m_{110}$ & 26.00$\pm$0.12 & 1 \\
$m_{127}$ & 24.52$\pm$0.12 & 1 \\
$m_{127}$ & 24.49$\pm$0.11 & 1 \\
$m_{127}$ & 24.36$\pm$0.09 & 1 \\
$J$ & $25.0^{+0.33}_{-0.53}$ & 4 \\
CH$_4$ cont & 23.2$\pm$0.2  & 1 \\
$H$ & $>22.7$\tablenotemark{a} & 5 \\
$K_s$ & $>18.6$\tablenotemark{a} & 6 \\
$W1$ & 17.82$\pm$0.33 & 5 \\
$W2$ & 14.02$\pm$0.05 & 5 \\
$[3.6]$ & 17.44$\pm$0.05 & 1 \\
$[3.6]$ & 17.30$\pm$0.05 & 1 \\
$[3.6]$ & 17.34$\pm$0.02 & 8 \\
$[3.6]$ & 17.28$\pm$0.02 & 8 \\
$[4.5]$ & 13.88$\pm$0.02 & 1 \\
$[4.5]$ & 13.90$\pm$0.02 & 1 \\
$[4.5]$ & 13.92$\pm$0.02 & 1 \\
$[4.5]$ & 13.93$\pm$0.02 & 1 \\
$[4.5]$ & 13.86$\pm$0.02 & 1 \\
$[4.5]$ & 13.82$\pm$0.02 & 1 \\
$[4.5]$ & 13.84$\pm$0.02 & 8 \\
$[4.5]$ & 13.86$\pm$0.02 & 1 \\
$[4.5]$ & 13.80$\pm$0.02 & 8 
\enddata
\tablecomments{All data are Vega magnitudes.
For bands with multiple measurements, the data are listed
in the order of the dates of observations from Table~\ref{tab:log} with
the exception of the second epoch in [3.6], for which a measurement is
not presented because of blending with another star.}
\tablenotetext{a}{S/N$<$3.}
\tablerefs{
(1) this work;
(2) \citet{kop14};
(3) \citet{bea14};
(4) \citet{fah14};
(5) \citet{wri14};
(6) VISTA Hemisphere Survey;
(7) \citet{luh14b};
(8) T. Esplin, in preparation.}
\end{deluxetable}

\begin{deluxetable}{ll}
\tabletypesize{\scriptsize}
\tablewidth{0pt}
\tablecaption{Proper Motion and Parallax of
WISE~J085510.83$-$071442.5\label{tab:prop}}
\tablehead{
\colhead{Parameter} & \colhead{Value} 
}
\startdata
$\pi$ & $0.449\pm$0.008$\arcsec$ \\
$\mu_{\alpha}$ cos $\delta$ & $-8.118\pm0.008\arcsec$~yr$^{-1}$  \\
$\mu_{\delta}$ & $0.680\pm0.007\arcsec$~yr$^{-1}$ 
\enddata
\end{deluxetable}

\clearpage

\end{document}